\documentclass[journal,twocolumn,10pt]{IEEEtran}

\usepackage{cite}
\usepackage{flushend}
\usepackage{booktabs}
\usepackage[pdftex]{graphicx}
\usepackage{amsmath,amssymb,amsfonts}
\usepackage{mathtools}
\usepackage{amsthm}
\usepackage{url}
\usepackage{xcolor}
\usepackage{enumitem}
\usepackage{xspace}
\usepackage{listings}

\usepackage[caption=false,font=footnotesize]{subfig}

\usepackage{standalone}
\usepackage{sansmath}
\usepackage{pgfplots}
\pgfplotsset{compat=newest}
\pgfplotsset{plot coordinates/math parser=false}
\usetikzlibrary{plotmarks,arrows.meta,positioning,calc}
\usepgfplotslibrary{patchplots}
\usepackage{grffile}

\definecolor{uclablue}{RGB}{39,116,174}
\definecolor{uclabluedarkest}{RGB}{0,59,92}
\definecolor{uclabluedarker}{RGB}{0,85,135}
\definecolor{uclabluelighter}{RGB}{139,184,232}
\definecolor{uclabluelightest}{RGB}{218,235,254}
\definecolor{uclagold}{RGB}{255,209,0}
\definecolor{uclagolddarker}{RGB}{255,199,44}
\definecolor{uclagolddarkest}{RGB}{255,184,28}
\definecolor{uclaviolet}{RGB}{130,55,165}
\definecolor{uclamagenta}{RGB}{255,0,165}
\definecolor{crimson}{RGB}{205,30,40}
\definecolor{violationfill}{RGB}{252,228,230}
\definecolor{outagefill}{RGB}{225,238,252}
\definecolor{degfill}{RGB}{218,235,254}

\newtheoremstyle{mydefinition}
{}
{}
{}
{0pt}
{\bfseries}
{.}
{ }
{\thmname{#1}\thmnumber{ #2}: \thmnote{#3}}

\theoremstyle{mydefinition}

\newtheoremstyle{myremark}
{}
{}
{}
{0pt}
{\bfseries}
{.}
{ }
{\thmname{#1}\thmnumber{ #2}: \thmnote{#3}}

\theoremstyle{myremark}

\newtheoremstyle{remarkshort}
{}
{}
{}
{0pt}
{\bfseries}
{.}
{ }
{\thmname{#1}\thmnumber{ #2}}

\theoremstyle{remarkshort}

\theoremstyle{remarkshort}

\newcommand{\subsecref}[1]{Subsection~\ref{#1}}
\newcommand{\tabref}[1]{Table~\ref{#1}}
\newcommand{\figref}[1]{Fig.~\ref{#1}}

\usepackage{xspace}

\let\originalleft\left
\let\originalright\right
\renewcommand{\left}{\mathopen{}\mathclose\bgroup\originalleft}
\renewcommand{\right}{\aftergroup\egroup\originalright}

\newcommand{\snr}{\ensuremath{\mathsf{SNR}}\xspace}
\newcommand{\sinr}{\ensuremath{\mathsf{SINR}}\xspace}
\newcommand{\inr}{\ensuremath{\mathsf{INR}}\xspace}

\usepackage{xspace}
\usepackage[acronym,nogroupskip,nonumberlist,nopostdot]{glossaries}
\makeglossaries
\usepackage{relsize}

\newacronym{epfd}{EPFD}{equivalent power flux-density}
\newacronym{fcc}{FCC}{Federal Communications Commission}
\newacronym{fss}{FSS}{fixed-satellite service}
\newacronym{gso}{GSO}{geostationary satellite orbit}
\newacronym{inr}{INR}{interference-to-noise ratio}
\newacronym{itu}{ITU}{International Telecommunication Union}
\newacronym{leo}{LEO}{low Earth orbit}
\newacronym{ngso}{NGSO}{non-geostationary satellite orbit}
\newacronym{rf}{RF}{radio frequency}
\newacronym{sinr}{SINR}{signal-to-interference-plus-noise ratio}
\newacronym{snr}{SNR}{signal-to-noise ratio}

\newcommand{\fcc}{\gls{fcc}\xspace}
\newcommand{\fss}{\gls{fss}\xspace}
\newcommand{\gso}{\gls{gso}\xspace}
\newcommand{\ginr}{\gls{inr}\xspace}
\newcommand{\itu}{\gls{itu}\xspace}
\newcommand{\leo}{\gls{leo}\xspace}
\newcommand{\ngso}{\gls{ngso}\xspace}
\newcommand{\rf}{\gls{rf}\xspace}
\newcommand{\gsinr}{\gls{sinr}\xspace}
\newcommand{\gsnr}{\gls{snr}\xspace}

\title{Spectrum Sharing Among Broadband LEO Mega-Constellations: Interference Regulations\\and Coexistence Mechanisms}

\author{Ian~P.~Roberts%
\thanks{I.~P.~Roberts is with the Wireless Lab, Department of Electrical and Computer Engineering, University of California, Los Angeles (UCLA), Los Angeles, CA 90095 USA (e-mail: ianroberts@ucla.edu).}%
}

\begin{document}

\maketitle

\begin{abstract}
Thousands of broadband LEO satellites are now in orbit with thousands more on the way, all operating under non-exclusive spectrum allocations across the Ku-, Ka-, and Q/V-bands. 
To govern this shared spectrum, regulators recently enacted protection constraints that cap instantaneous interference levels and time-averaged throughput degradation. 
In this article, we review the governing FCC and ITU regulatory frameworks and dissect this interference environment. 
We examine practical non-cooperative coexistence mechanisms, highlighting how dynamic satellite selection and passive ground sensing circumvent the physical limits of beamforming alone.
We then evaluate potential cooperative coordination architectures spanning centralized registries and bilateral interfaces.
Finally, we analyze key open challenges: scaling beyond two networks, mitigating feeder link contention from direct-to-cell systems, rerouting traffic over optical inter-satellite links, and modernizing international sharing policies.
\end{abstract}


\glsresetall

\section{Introduction} \label{sec:introduction}

More than 10{,}000 commercial communications satellites currently operate in \leo, with another 50{,}000+ slated to launch over the coming decade.
Orbiting primarily at 500--1{,}200~km overhead, these spacecraft cut one-way free-space propagation delay to a few milliseconds and significantly reduce path loss compared to traditional \gso systems, positioning \leo satellite networks to complement or even compete with terrestrial networks in certain markets.

Unlike terrestrial cellular networks, which obtain exclusive spectrum licenses at auction, broadband \leo satellite providers operate under \textit{non-exclusive} \fss allocations governed by the \itu and national regulators such as the United States \fcc.
As a result of this non-exclusivity, multiple \leo constellations may serve overlapping geographic regions at the same frequencies.
This causes spectrum contention that is largely divided between two operational links, as mapped in \figref{fig:spectrum_sharing_fundamentals}(a):
\begin{itemize}
    \item \textbf{\emph{User service links}} connect satellites directly to user terminals, operating primarily across Ku-band (10.7--12.7~GHz) and Ka-band (17.7--20.2~GHz).
    \item \textbf{\emph{Gateway feeder links}} connect satellites to terrestrial fiber backbones, utilizing Ka-band and increasingly Q/V-band (37.5--51.4~GHz) for high-capacity backhaul.
\end{itemize}
Contention arises between competing user downlinks, between gateway feeder links, and where user and feeder allocations intersect. In the Ka-band (17.7--20.2~GHz), for instance, Amazon Leo (formerly Kuiper) operates user downlinks over the same frequencies that Starlink and OneWeb use for high-capacity gateway downlinks \cite{amazon_kuiper_filing, fcc_starlink_gen2}.
Mitigating this interference is central to enabling sustainable spectrum coexistence among \leo mega-constellations.

\subsection{Satellite Operation and Interference Dynamics} \label{subsec:satellite_operation_interference_dynamics}
\begin{figure*}[!t]
    \centering
    \includegraphics[width=\textwidth]{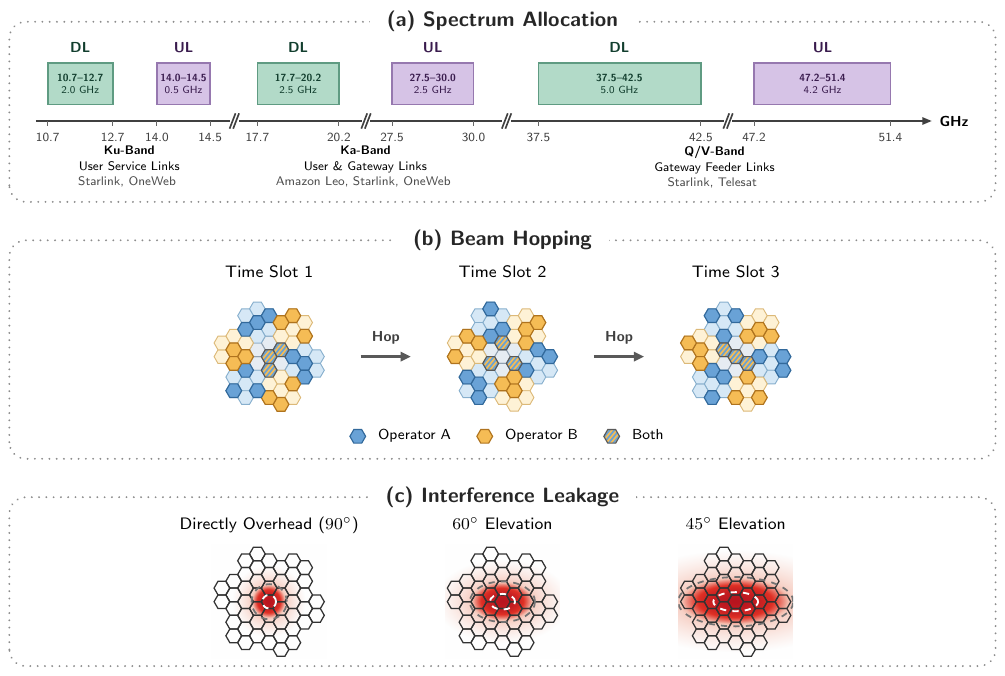}
    \caption{Spectrum sharing fundamentals in \leo mega-constellations: (a) shared Ku-, Ka-, and Q/V-band allocations; (b) dynamic spot beam hopping across cells; and (c) elevation-dependent leakage into adjacent cells (based on the ITU-R S.1528 beam mask and 20~km cells).}
    \label{fig:spectrum_sharing_fundamentals}
\end{figure*}

To maximize capacity across millions of subscribers, modern \leo networks partition their wideband spectrum allocations into smaller channels.
For example, Starlink reportedly divides its 2~GHz Ku-band spectrum into eight 250~MHz channels \cite{humphreys_2023_starlink_signal}.
High-gain transmissions are directed toward ground cells, each on the order of 20~km in diameter, using active antenna arrays capable of generating dozens of spot beams on each frequency channel.
Satellites then use \textit{beam hopping} (or \textit{time-division multiplexing}) to dynamically steer these beams across a greater number of cells, dwelling on each cell for millisecond-scale bursts, as illustrated in \figref{fig:spectrum_sharing_fundamentals}(b).
As a representative example, a single satellite may form 16 spot beams on three different frequency channels, totaling 48 simultaneous beams; a beam hopping factor of three would thus allow that satellite to serve 144 ground cells on its own~\cite{amazon_kuiper_filing, fcc_starlink_gen2}.
As satellites traverse their orbits, handovers periodically reassign each satellite to a new cluster of ground cells, observed to happen every 15~seconds~\cite{qin_2026_starlink_timing, mohan_2024_starlink_performance}.

Interference arises when two satellites from competing constellations illuminate overlapping or neighboring ground cells on the same frequency.
While directional user terminals within these cells provide isolation when satellites maintain wide angular separation, high interference can occur when spacecraft in intersecting orbital planes pass in close angular proximity~\cite{braun_2019_dyspan}.
This worsens at lower elevation angles, since elongated beam footprints increase leakage into adjacent cells (\figref{fig:spectrum_sharing_fundamentals}(c)).
Because satellites travel at roughly 7.5~km/s, this interference geometry evolves rapidly across dozens of visible spacecraft and hundreds of spot beams.
In this article, we examine this coexistence environment by detailing the governing regulatory frameworks, evaluating non-cooperative mitigation mechanisms, examining cooperative coordination architectures, and outlining open technical and policy challenges.

\section{Dissecting FCC and ITU Regulations} \label{sec:dissecting_fcc_itu_regulations}
\subsection{FCC Processing Rounds and Spectrum Sharing} \label{subsec:fcc_processing_rounds_spectrum_sharing}
To grant non-exclusive spectrum licenses to \leo operators, the \fcc relies on a procedural framework built on the concept of a \textit{processing round}.
When an operator applies to launch a new \ngso constellation, the \fcc initiates a public proceeding inviting competing entities to file mutually exclusive or overlapping applications by a specified cut-off date.
All qualified applicants authorized within the same processing round share the assigned frequencies on a co-equal basis, meaning no single system holds spectrum priority over another authorized in that round.

The \fcc ties spectrum rights to specific satellite \textit{constellations} rather than corporate entities, placing distinct constellation generations from the same operator into separate priority tiers.
For example, SpaceX Starlink Gen1, Eutelsat OneWeb, and Telesat secured authorization through the 2016--2017 Ku/Ka-band round, whereas Amazon Leo and SpaceX Starlink Gen2 participated in subsequent rounds.
\tabref{tab:fcc_timeline} summarizes how this processing-round framework and its associated sharing rules have evolved over the past decade.

\begin{table*}[!t]
    \centering
    \caption{Evolution of \fcc NGSO FSS spectrum-sharing rules and processing rounds.}
    \label{tab:fcc_timeline}
    \footnotesize
    \begin{tabular}{@{}ll@{}}
        \toprule
        \textbf{Year} & \textbf{Development} \\
        \midrule
        1997 & \fcc establishes the foundational Ka-band \fss framework underlying today's \ngso rules. \\
        2016--2017 & First Ku/Ka- and V-band rounds authorize SpaceX Gen1, OneWeb, and Telesat under equal band-splitting ($-12.2$~dB INR). \\
        2020 & Second Ku/Ka-band round opens (Amazon Leo); SpaceX petitions the \fcc to update band-splitting to protect earlier rounds. \\
        2021 & Second V-band round opens; \fcc proposes limiting equal band-splitting to same-round systems and adding inter-round protection. \\
        2023 & \fcc 23-29 splits sharing into intra-round parity and inter-round protection, adding a ten-year sunset \cite{fcc_23_29}. \\
        2024 & \fcc 24-117 sets inter-round thresholds at $3\%$ throughput degradation and $0.4\%$ link unavailability \cite{fcc_24_117}. \\
        2026 & \fcc 26-26 proposes replacing \ngso-to-\gso EPFD limits with $3\%$ throughput degradation and $0.1\%$ link unavailability thresholds \cite{fcc_26_26}. \\
        \bottomrule
    \end{tabular}
\end{table*}

To govern spectrum sharing between constellations (both within the same processing round and across different rounds), the \fcc codified explicit sharing rules in 2023 under 47 CFR \S~25.261, creating two distinct operational regimes~\cite{fcc_23_29, fcc_24_117}:
\begin{itemize}
    \item \textbf{\emph{Intra-Round Sharing (Co-Equal Parity):}}
    Operators authorized in the same processing round must coordinate in good faith.
    If they cannot reach an agreement, a default rule mandates equal band-splitting whenever interference exceeds allowable thresholds.

    \item \textbf{\emph{Inter-Round Sharing (Incumbent Protection):}}
    Constellations authorized in earlier processing rounds hold priority over later entrants, meaning newer systems must protect incumbent networks from excessive throughput degradation or link outages.
\end{itemize}

\subsection{Intra-Round Interference Protection Constraints} \label{subsec:intra_round_interference_constraints}
To put these intra-round and inter-round rules into practice, regulators establish quantitative measures that define when interference becomes excessive.
For intra-round co-equal sharing, the criterion is based on instantaneous receiver noise degradation.
Under 47 CFR \S~25.261(c), mandatory band-splitting applies whenever the aggregate interference power from a competing system causes an increase in receiver equivalent noise temperature exceeding $6\%$ \cite{fcc_23_29, fcc_24_117}.
This $6\%$ noise temperature increase translates directly into an \ginr limit of $-12.2$~dB, meaning aggregate interference inflicted on a receiver must remain below roughly one-seventeenth of its noise floor.

This noise-based protection constraint has long been used in satellite coordination because it normalizes interference across systems with differing bandwidths, as both thermal noise and wideband interference scale (roughly) linearly with bandwidth.
Furthermore, because noise degradation is independent of the user's desired signal strength, an operator can evaluate its own compliance based solely on the interference power it inflicts, given an estimate of the victim receiver's nominal noise floor.
By contrast, a \gsinr-based protection constraint would depend on the victim's nominal \gsnr, which is typically proprietary information. 

\subsection{Inter-Round Interference Protection Constraints} \label{subsec:inter_round_interference_constraints}

While an instantaneous $-12.2$~dB threshold indicates when interference becomes excessively high, it does not directly reflect its impact on system throughput.
In light of this, to govern inter-round protection, the \fcc modernized its framework in 2024 under 47 CFR \S~25.261(d) by adopting two performance-based criteria evaluated over an orbital simulation period $T$ \cite{fcc_23_29, fcc_24_117}:
\begin{enumerate}[leftmargin=*,label=(\roman*)]
    \item \textbf{\emph{Throughput Degradation ($\le 3\%$):}}
    The time-weighted reduction in throughput on an incumbent link caused by an entrant's transmissions must not exceed $3\%$, i.e., 
    \begin{equation} \label{eq:throughput_degradation}
        1 - \frac{\int_T R(\sinr(t)) \, \mathrm{d}t}{\int_T R(\snr(t)) \, \mathrm{d}t} \le 0.03,
    \end{equation}
    where $R(\cdot)$ maps instantaneous link quality to throughput according to the incumbent's modulation and coding tables, $\snr(t)$ is the nominal \gsnr\footnote{Throughout this article, $\mathsf{SNR}$ denotes the carrier-to-noise
    	ratio at the user terminal receiver, measured over the carrier's noise
    	bandwidth, i.e., the quantity the \fcc writes as $C/N$.} at time $t$, and $\sinr(t) = \frac{\snr(t)}{1+\inr(t)}$ accounts for co-channel interference $\inr(t)$.
    To qualify for this protection, an incumbent link must demonstrate baseline availability of at least $99\%$ at $\snr = 0$~dB (absent any interference), preventing operators from claiming protection for links deployed without adequate reliability.

    \item \textbf{\emph{Link Unavailability Increase ($\le 0.4\%$):}}
    The absolute increase in the percentage of time that the incumbent link quality drops below a fixed $\snr = 0$~dB benchmark must not exceed $0.4\%$, corresponding to an additional outage allowance of roughly 35~hours per year.
    Benchmarking unavailability at a fixed \gsnr rather than a demodulation threshold ensures an entrant can evaluate compliance using public link baselines without requiring access to an incumbent's proprietary modulation and coding tables.
\end{enumerate}
\begin{figure}[!t]
    \centering
    \includegraphics[width=\columnwidth]{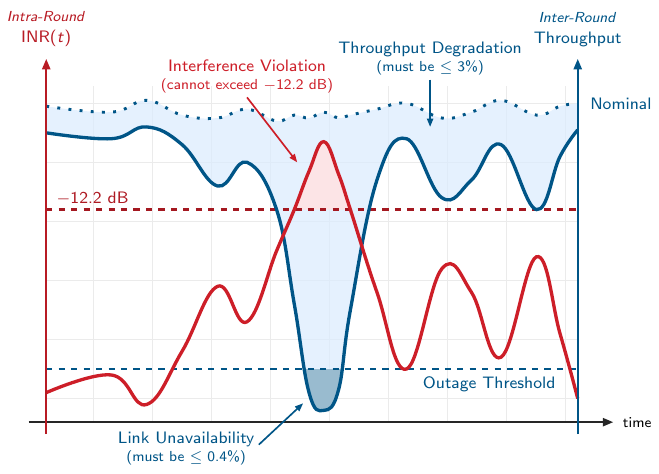}
    \caption{An illustration of the \fcc coexistence criteria under 47 CFR \S~25.261~\cite{fcc_23_29,fcc_24_117}. An instantaneous $-12.2$~dB \ginr threshold among intra-round, co-equal constellations versus time-averaged $3\%$ throughput degradation and $0.4\%$ link unavailability limits among inter-round constellations.}
    \label{fig:coexistence_regimes}
\end{figure}

\figref{fig:coexistence_regimes} illustrates this dual regulatory structure, contrasting instantaneous interference thresholds for intra-round sharing with time-averaged performance degradation for inter-round protection.\footnote{The \fcc extended this performance-based philosophy to \gso networks in April 2026 under FCC~26-26~\cite{fcc_26_26}, proposing to replace static EPFD limits with a $3\%$ throughput degradation and $0.1\%$ unavailability standard.}
Under \fcc rules, both criteria apply on a \textit{per-link and per-interfering-system} basis, currently evaluated strictly \textit{ex ante} through orbital simulations submitted during licensing rather than through real-time measurements.
This per-system standard prevents an entrant from having its operational rights curtailed by third-party transmissions, but it creates a cumulative loophole: two later-round constellations could each degrade an incumbent link by $3\%$, inflicting an aggregate reduction exceeding $6\%$ while each entrant remains individually compliant (an issue revisited in \subsecref{subsec:multi_constellation_scaling}).

Finally, the \fcc enforces a ten-year sunset provision under \S~25.261(e): ten years after the first authorization is granted in an entrant's processing round, those entrant systems are no longer required to protect earlier incumbents, reverting instead to co-equal band-splitting.
Starting the clock on the entrant round's grant date (rather than the incumbent's earlier licensing date) guarantees that incumbents receive a full ten years of protection while the entrant deploys its constellation across its statutory build-out milestones ($50\%$ by year six, $100\%$ by year nine under \S~25.164).

\subsection{ITU Regulations} \label{subsec:itu_regulations}
While these \fcc regulations establish clear domestic rules within United States jurisdiction, mega-constellations largely operate globally under the international treaty framework of the \itu.
Unlike the \fcc's prescriptive framework, the \itu does not formally quantify excessive interference for coexistence between \ngso constellations.
The \itu defines ``harmful interference'' strictly qualitatively as interference that endangers safety services or seriously degrades, obstructs, or repeatedly interrupts transmissions~\cite{itu_radio_regulations}.
To protect \textit{\gso} networks, the \itu traditionally enforces \gls{epfd} limits under Article~22 evaluated using standardized software tools, while its underlying framework offers advisory \ginr guidance broadly consistent with the \fcc's $-12.2$~dB threshold for co-primary services, though the exact allowance varies with the class and number of interfering networks.
For coexistence among competing \ngso constellations, however, the \itu framework hinges entirely on date-of-receipt filing priority under Article~9~\cite{itu_radio_regulations}.
Under this regime, later-filed constellations bear the burden of negotiating bilateral coordination agreements or ceasing operations to avoid harmful interference, with no default band-splitting remedies, throughput degradation constraints, or sunset provisions comparable to those under the \fcc~\cite{fcc_23_29, fcc_24_117}.
Historically, this first-come priority incentivized operators to submit speculative, early-stage filings to secure international priority dates.
To curb these ``paper constellations,'' the \itu established a milestone-based deployment framework under Resolution~35: following an initial period to bring their assignments into use, operators must deploy $10\%$ of their authorized constellation within two years, $50\%$ within five years, and $100\%$ within seven years, or forfeit unbuilt capacity~\cite{itu_wrc19_res35}.

This divergence creates significant operational friction for global operators.
Two constellations sharing spectrum under co-equal band-splitting rules within the United States must revert to strict first-come priority over international waters and foreign territories.
To operate globally, constellations must therefore maintain region-dependent transmission and scheduling rules that switch dynamically across national borders.
Until international frameworks harmonize these regulatory procedures, constellations must rely on unilateral operational mechanisms to prevent interference across jurisdictions.

\section{Non-Cooperative Coexistence Mechanisms} \label{sec:non_cooperative_coexistence_mechanisms}

Regardless of jurisdiction, constellations must avoid inflicting excessive interference onto co-frequency receivers while maximizing radio resource utilization.
Because competing commercial operators are disincentivized from sharing proprietary traffic demands and lack standardized interfaces for real-time coordination, constellations must initially rely on \textit{non-cooperative} techniques to prevent interference.
This section explores such techniques across frequency, time, power, and spatial domains that facilitate coexistence without requiring coordination between operators.

\subsection{Dynamic Resource Allocation} \label{subsec:dynamic_resource_allocation}
When unoccupied spectrum exists, the simplest defense against interference is to avoid transmitting on the same frequencies as a competing constellation.
In the frequency domain, an operator can reallocate active spot beams into uncontested sub-channels, creating a temporary spectral notch over the affected ground cell.
Complementing this in the time domain, dynamic beam hopping allows an operator to adjust dwell times and illumination sequences so that transmissions targeting adjacent cells are interleaved rather than radiated simultaneously.
When spectral and temporal avoidance are insufficient, adaptive power control can be used to throttle spot beam transmit power to keep interference below the $-12.2$~dB threshold.

\begin{figure*}[!t]
    \centering
    \includegraphics[width=\textwidth]{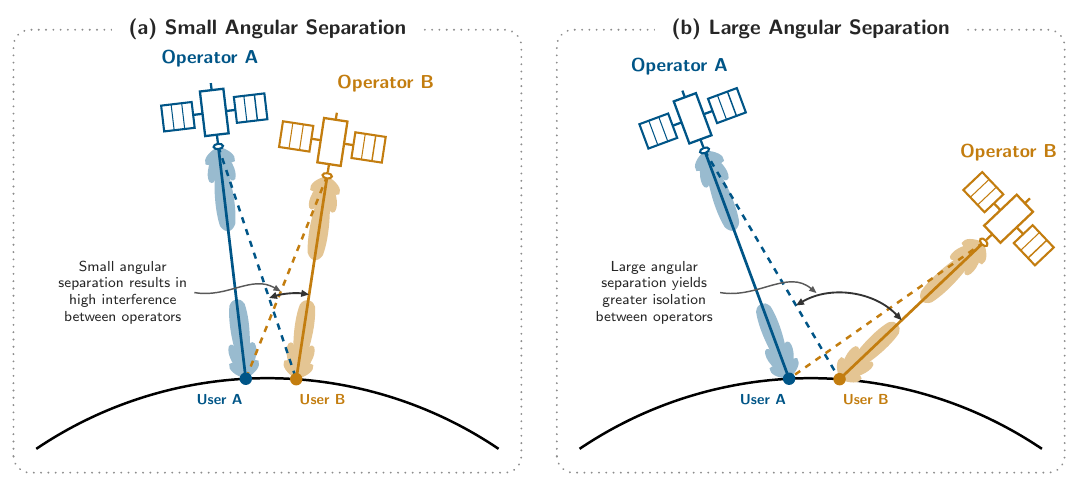}
    \caption{Angular separation between competing \leo constellations: (a) small angular separation leads to main lobe alignment and thus high interference, and (b) large angular separation provides isolation between the two satellites.}
    \label{fig:angular_separation}
\end{figure*}

While these dynamic resource allocation techniques can be effective, they come at the cost of reduced capacity.
Dividing spectrum into narrower sub-channels reduces user data rates and introduces guard-band overhead.
Similarly, shortening beam dwell times diminishes aggregate capacity in high-demand service cells.
Reducing transmit power lowers received SNR on the operator's own links, degrading throughput and risking link outages.

\subsection{Phased-Array Beamforming and Nulling} \label{subsec:phased_array_beamforming_nulling}

When frequency, time, or power adjustments fall short, the remaining recourse is spatial isolation.
Today's \leo satellites employ phased arrays with hundreds of antenna elements to form highly directional spot beams.
By dynamically adjusting beamforming weights, an operator can suppress sidelobes or steer nulls toward competing ground users to reduce interference.
Although the regulatory burden of protection rests primarily on the satellite, a ground user could similarly taper its receive sidelobes or steer nulls toward competing satellites to further reduce interference, assuming it has the knowledge and ability to do so.

In practice, synthesizing beams and nulls with high precision on orbit is constrained by hardware and environmental limits.
First, \rf front-end non-idealities and array miscalibration distort beam patterns, a challenge exacerbated by thermal variations that cause time-varying amplitude and phase drift across array elements.
Without continuous array calibration, these drifts degrade achievable null depths and sidelobe suppression compared to idealized models.
Second, attitude estimation and pointing errors shift the antenna orientation relative to ground coordinates; because nulls are extremely narrow, an error of even a fraction of a degree can displace a null away from a victim receiver, sharply increasing interference.
Third, forming nulls or tapering sidelobes consumes spatial degrees of freedom, which widens main lobes, lowers peak gain, and limits the total number of simultaneous spot beams a satellite can generate.
A user terminal faces similar constraints: while it can track a serving satellite with its main lobe, it has limited knowledge of its exact orientation and likely lacks the high-fidelity calibration needed to reliably place nulls toward moving satellite interferers.
Compounding this, cost-constrained user terminals typically use analog beamformers with coarse phase shifters, restricting both null depth and sidelobe suppression.

\subsection{Interference Mitigation Through Satellite Selection} \label{subsec:satellite_selection}
Even with ideal beamforming, spatial isolation fails when a satellite from a competing constellation aligns along the line of sight between an operator's satellite and its ground users.
During these in-line events, the interfering spacecraft and the serving satellite appear in nearly the same angular direction from the ground's perspective.
Consequently, both the desired signal and interference enter through the user terminal's high-gain main lobe (\figref{fig:angular_separation}(a)), preventing it from suppressing the interferer without also suppressing its desired link.

\begin{figure*}[!t]
    \centering
    \includegraphics[width=\textwidth]{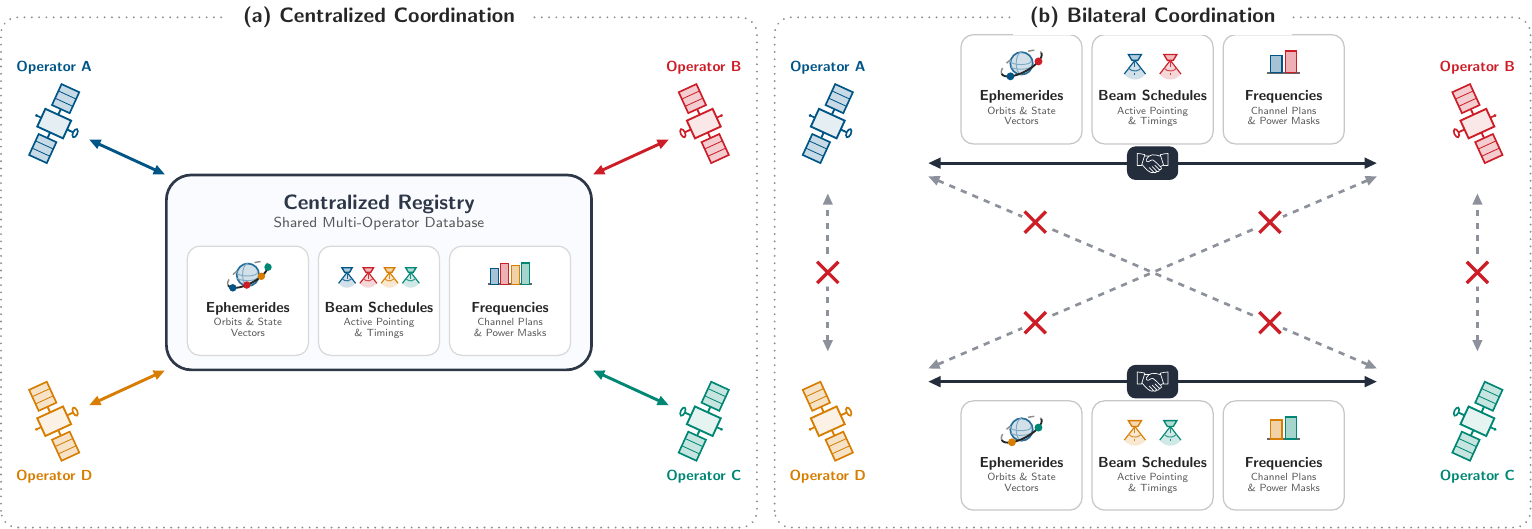}
    \caption{Cooperative spectrum coordination architectures among \leo constellations: (a) centralized coordination through a shared database and (b) bilateral coordination via private inter-operator agreements.}
    \label{fig:coordination_architectures}
\end{figure*}

Outside these in-line events, spatial isolation without beamforming relies on antenna pattern roll-off at both ends of the link.
At the satellite, reference radiation masks such as ITU-R S.1528~\cite{itu_r_s1528} exhibit a near-sidelobe plateau that limits off-axis attenuation to approximately $20$~dB for ground users located 25--65~km from beam center (assuming a 550~km orbital altitude).
Achieving $40$~dB of satellite-side suppression requires nearly 500~km of geographic separation on the ground.
When competing constellations illuminate adjacent service cells, satellite transmit patterns alone cannot prevent excessive interference.
Downlink isolation therefore hinges on the user terminal's receive pattern and its angular separation from interfering satellites (\figref{fig:angular_separation}(b)).
Under user terminal reference models such as ITU-R S.1428~\cite{itu_r_s1428} (e.g., a 0.5-m aperture at 11.7~GHz with $33.5$~dBi gain), sidelobe suppression plateaus at $36.4$~dB beyond $48^\circ$ off-boresight.
Suppressing aggregate interference below $-12.2$~dB requires an angular separation of at least $8.1^\circ$ for a single interfering satellite, $15.5^\circ$ for five, and $27.0^\circ$ for twenty, assuming equal effective isotropic radiated power density, comparable slant ranges, and a user terminal located at the interfering beam's $-3$~dB contour on an $8$~dB link.
On stronger $12$--$16$~dB links, these angular requirements widen substantially; on a $16$~dB link with 14 or more visible spacecraft, the required isolation exceeds the user terminal's $36.4$~dB sidelobe floor entirely.
Because beamforming by a single satellite cannot overcome these physical sidelobe limits during close angular alignments, operators must exploit spatial diversity across the constellation to achieve sufficient isolation.

In dense modern \leo networks, dozens of satellites (often 20 or more today) are typically visible above the minimum elevation angle at any given time.
Because satellite orbits are deterministic and publicly available, an operator can anticipate when a satellite of another constellation will align with one of its own.
The operator can then use this knowledge to proactively hand over ground cells to an alternate satellite that provides sufficient angular separation, protecting unintended ground receivers while sustaining high throughput on its own downlinks~\cite{kim_2026_satellite_selection}.
Steering clear of a competing satellite is in fact mutually beneficial, since it also shields the operator's own ground users from incoming interference.
Recent work~\cite{kim_2026_satellite_selection} demonstrates that this satellite selection strategy alone, without adaptive array null steering, can satisfy regulatory interference thresholds across large-scale networks while preserving over $95\%$ of link capacity.

\subsection{Learning Operational Policies from Ground Observations} \label{subsec:learning_operational_policies_ground_observations}
Proactive satellite selection is most effective when an operator knows where a competing constellation's beams are directed in real time. 
However, active beam schedules and user locations remain strictly proprietary.
Without direct coordination, an operator must fall back on conservative link budgets derived from public regulatory filings, forcing it to assume worst-case interference and unnecessarily sacrifice network capacity.

Rather than rely on static worst-case assumptions, an operator may use passive ground observations to directly infer a competing constellation's activity patterns and their underlying policies.
Even without demodulating user traffic, an operator can monitor downlink transmissions using its own user terminals or dedicated ground sensors to detect spectral occupancy and beam activity.
Correlating these observations with the known orbits of overhead satellites may then allow an operator to learn a competing network's scheduling policy and predict its beam activity in advance~\cite{kim_2026_satellite_selection}.
This foresight would enable proactive satellite handovers and beam steering to mitigate interference before it occurs.
Ground-based measurements could also provide regulators with an empirical mechanism to verify compliance in orbit, addressing the limitations of relying strictly on static \textit{ex ante} orbital simulations.

\begin{figure*}[!t]
    \centering
    \includegraphics[width=\textwidth]{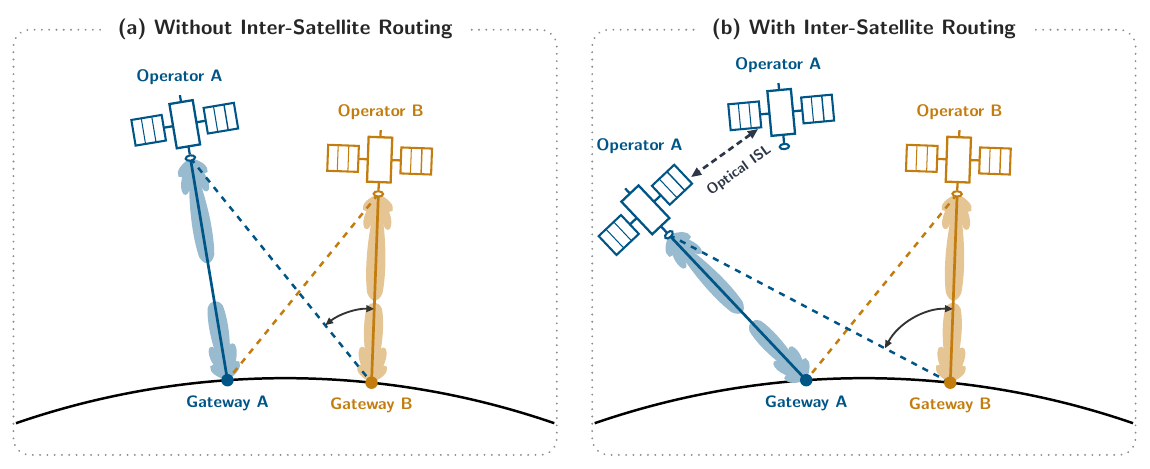}
    \caption{Inter-satellite routing for spectrum coexistence: (a) without inter-satellite links (shown as ISL), local gateway links suffer high interference during in-line events, and (b) with inter-satellite routing, traffic is relayed through alternate satellites that provide greater angular separation.}
    \label{fig:isl_routing}
\end{figure*}

\section{Cooperative Spectrum Coordination} \label{sec:cooperative_spectrum_coordination}
Non-cooperative satellite selection and beamforming allow operators to mitigate interference independently, but doing so with incomplete or inaccurate information risks inadvertently violating protection constraints or unnecessarily sacrificing capacity.
Direct coordination between operators eliminates this uncertainty, enabling constellations to proactively manage shared spectrum and prevent excessive interference events.

\subsection{Extending Orbital Safety Frameworks to Spectrum Sharing} \label{subsec:extending_orbital_safety_frameworks}
Historically, satellite spectrum coordination has relied on manual negotiations and bilateral agreements between operators, designed primarily for \gso orbital arc allocations.
In dense \leo mega-constellations, however, thousands of spacecraft traverse the sky at high velocities while dynamically steering spot beams, rendering manual or static coordination more challenging.

To manage the dynamics of these constellations, the satellite communications industry could adapt the automated coordination frameworks established for orbital space safety.
For physical collision avoidance, operators routinely share high-precision orbital ephemerides and standardized data messages that forecast close approaches days in advance.
Extending automated data exchange to spectrum sharing would allow operators to anticipate spatial alignments and spectrum contention hours in advance.
By exchanging predictive ephemerides, coverage footprints, and planned beam schedules, constellation schedulers could identify upcoming conflicts and adjust satellite-to-cell assignments or frequency allocations before harmful interference occurs.

However, sharing operational schedules naturally introduces strategic and game-theoretic tensions.
When an upcoming interference event is identified, deciding which operator yields---e.g., by handing over to an alternate satellite, reducing transmit power, or vacating a sub-band---creates competing incentives.
Under co-equal intra-round sharing, neither operator holds spectrum priority, creating a deadlock over who yields unless mandatory band-splitting takes effect.
By contrast, inter-round rules place the legal burden strictly on the later entrant to protect the incumbent.
In either case, without standardized coordination protocols, operators remain reluctant to sacrifice their own link capacity or reveal proprietary traffic demands to a competitor.

\subsection{Centralized Registries vs.~Bilateral Interfaces} \label{subsec:centralized_registries_vs_bilateral_interfaces}
This exchange of operational data could occur through either a centralized or bilateral architecture, as illustrated in \figref{fig:coordination_architectures}.
In a centralized architecture (\figref{fig:coordination_architectures}(a)), all constellations regularly submit their orbital ephemerides, antenna models, and planned beam schedules to a shared registry.
This mirrors both space safety clearinghouses (such as Space-Track and the Space Data Association) and dynamic terrestrial spectrum databases (such as 6~GHz Wi-Fi Automated Frequency Coordination and the 3.5~GHz Spectrum Access System).
While updating a centralized registry on millisecond beam-hopping timescales is impractical, it could realistically be updated at regular satellite-to-cell reassignment intervals (e.g., every 15~seconds) or populated with planned schedules hours in advance.
The registry would function primarily as an information clearinghouse to forecast potential interference events, leaving mitigation and arbitration to individual operators themselves.
In a bilateral model (\figref{fig:coordination_architectures}(b)), individual pairs of operators establish private interfaces directly between one another.
While these bilateral arrangements protect proprietary commercial data from competitors, a centralized registry offers transparency and scales more effectively to a greater number of constellations.

\section{Open Challenges and Future Directions} \label{sec:open_challenges_future_directions}

\subsection{Multi-Constellation Scaling Bottlenecks} \label{subsec:multi_constellation_scaling}
As additional mega-constellations enter the market, operational scenarios will routinely involve three or more networks competing for the same spectrum.
Although two-operator coexistence can be resolved through strategic satellite selection or band-splitting, these unilateral mechanisms face scaling limits across both frequency and spatial domains when three or more networks occupy overlapping orbital shells.

The default regulatory remedy of mandatory equal band-splitting, for instance, degrades rapidly as the number of operators grows.
While dividing spectrum between two operators already halves available bandwidth, splitting it among three or four networks further fragments allocations into narrow sub-channels, reducing data rates and compounding guard-band overhead~\cite{ozturk_2025_spectral_efficiency}.
Beamforming faces comparable constraints, as synthesizing simultaneous nulls toward receivers across multiple independent networks exhausts array degrees of freedom.
Satellite selection likewise reaches geometric limits. As more constellations populate the sky, maintaining angular clearance from multiple competing links leaves few viable pointing directions.
Managing these crowded multi-operator environments will require frameworks that incentivize multi-party coordination to prevent the steep capacity losses caused by severe spectrum fragmentation or uncoordinated interference.

\subsection{Feeder Link Contention from Direct-to-Cell Systems} \label{subsec:feeder_link_contention_direct_to_cell}
The emergence of direct-to-cell constellations introduces an additional source of spectrum contention.
Unlike broadband systems that serve user terminals over Ku- and Ka-bands, direct-to-cell satellites communicate directly with unmodified smartphones using frequencies below 3~GHz on user links.
To backhaul this traffic, however, their high-throughput feeder links rely heavily on the same Ka- and Q/V-band allocations used by broadband networks.
Amazon Leo, for instance, filed in July 2026 for a 5{,}105-satellite direct-to-cell constellation operating mobile-satellite service links in the L- and S-bands (1.6/2.4~GHz) while routing backhaul through its existing Ka-band \fss gateways.
Backhauling traffic for thousands of direct-to-cell satellites substantially increases gateway transmission activity across these shared feeder bands, escalating contention with existing broadband systems.

Fortunately, unlike user terminals, whose locations are proprietary, gateways are fixed installations with publicly registered coordinates.
This permanence simplifies spatial mitigation, since transmitting satellites can steer pattern nulls directly toward known gateway coordinates without needing to infer their locations.
Because orbital trajectories and gateway coordinates are both deterministic, operators can reliably forecast in-line events and proactively hand over feeder links to alternate satellites to prevent interference.

\subsection{Inter-Satellite Routing} \label{subsec:inter_satellite_routing}
Modern \leo constellations increasingly deploy optical inter-satellite links, allowing spacecraft to route data directly to one another in orbit at extremely high data rates.
While optical links primarily serve to connect remote regions lacking terrestrial infrastructure, inter-satellite routing also provides distinct advantages in terms of spectrum coexistence.

First, inter-satellite links provide operational agility when local feeder links experience contention, as illustrated in \figref{fig:isl_routing}.
When two satellites pass in close angular proximity over neighboring gateways (\figref{fig:isl_routing}(a)), they inflict excessive interference on one another.
To mitigate this, an operator can relay feeder traffic over optical links to an alternate satellite that illuminates the same gateway from a wider, uncontested angle (\figref{fig:isl_routing}(b)).
In essence, this extends satellite selection from user downlinks to gateway feeder links.
In remote regions or during severe congestion, inter-satellite routing can similarly offload traffic entirely to distant gateways thousands of kilometers away, bypassing contested ground infrastructure altogether.

Second, inter-satellite routing directly relieves the feeder link congestion discussed in \subsecref{subsec:feeder_link_contention_direct_to_cell}.
By relaying data between satellites, constellations can aggregate traffic into fewer physical gateways, though this consolidation naturally has its limits.
Even so, reducing the number of high-gain gateways operating across shared Ka- and Q/V-bands eases coexistence with emerging direct-to-cell systems.

\subsection{Global Policy and Standardized Protocols} \label{subsec:global_policy_standardized_protocols}
Deploying automated spectrum coordination at scale requires international standards developed through bodies such as the \itu and 3GPP.
To coordinate effectively without exposing proprietary scheduling algorithms or traffic demands, regulatory standards should establish frameworks for exchanging non-sensitive operational data, such as coverage footprints and satellite assignments, while keeping dynamic beam-hopping sequences and user traffic private.
Beyond coordination protocols, global regulatory frameworks must also modernize sharing rules.
Following the \fcc's transition to performance-based metrics, establishing comparable criteria within the \itu is needed to accommodate coexistence mechanisms and support spectrum sharing as mega-constellations expand.

\bibliographystyle{bibtex/IEEEtran}
{\small \bibliography{IEEEabrv,myrefs}}

\end{document}